\documentclass[fleqn,usenatbib,useAMS]{mnras}

\usepackage{newtxtext,newtxmath}
\usepackage[T1]{fontenc}
\usepackage{ae,aecompl}

\usepackage{graphicx}	% Including figure files
\usepackage[caption=false]{subfig} %  for subfigures environments
\usepackage{amsmath}	% Advanced maths commands
\usepackage{amssymb}	% Extra maths symbols
\usepackage[dvipsnames]{xcolor}

\newcommand{\argdf}{\texttt{ARGDF}}
\newcommand{\archain}{\texttt{AR-CHAIN }}

\title[Binary stars hosting planets approaching Sgr A*]{The Fate of Binary Stars Hosting Planets upon Interaction with\\ Sgr A* Black hole}

\author[R. Capuzzo-Dolcetta and N. Davari]{
R. Capuzzo-Dolcetta \thanks{E-mail: roberto.capuzzodolcetta@uniroma1.it} 
and
N. Davari \thanks{E-mail: nazanin.davari@uniroma1.it}
\\
Dep. of Physics, Sapienza, Univ. of Rome, P.le A. Moro 5, I00185 Rome, Italy\\
}

\pubyear{2018}

\begin{document}
\label{firstpage}
\pagerange{\pageref{firstpage}--\pageref{lastpage}}
\maketitle

\begin{abstract}
Our Galaxy hosts a very massive object at its centre, often referred to as the supermassive black hole Sgr A*. Its gravitational tidal field is so intense that can strip apart a binary star passing its vicinity and accelerate one of the components of the binary as hypervelocity star (HVS) and grab the other star as S-star.
Taken into consideration that many binary star systems are known to host planets, in this paper we aim to broaden the study of the close interaction of binary stars and their planetary systems with Sgr A* massive object. Results are obtained via a high precision $N-$body code including post-Newtonian approximation.  
We quantify the likelihood of capture and ejection of stars and planets after interaction with Sgr A*, finding that the fraction of stars captured around it is about three times that of the planets  ($\sim 49.4$ \% versus $\sim 14.5$ \%) and the fraction of hypervelocity planet (HVP) ejection is about twice HVSs ($\sim 21.7$ \% versus $\sim 9.0$ \%). The actual possibility of observational counterparts deserves further investigation.\par
\end{abstract}


%%%%%%%%%%%%%%%%%%%%%%%%%%%%%%%%%%%%%%%%%%%%%%%%%%

%%%%%%%%%%%%%%%%% BODY OF PAPER %%%%%%%%%%%%%%%%%%

\section{Introduction}

The presence of a compact massive object (CMO) with the mass of $M_{\bullet} \sim 4\times 10^{6} M_\odot$ is ascertained in the center of our Galaxy, although the same evidence of an event horizon around a supermassive black hole (SMBH) like for the giant elliptical M 87, has not yet been given \citep{2019ApJ...875L...1E}. The interaction of such CMO with passing-by stars and/or whole stellar systems may have significant consequences. As first hypothesized by Hills (\citeyear{1988Natur.331..687H}), a binary star passing close enough to a massive black hole (MBH) can be broken by its tidal field and expel one star as a hypervelocity star (HVS) while the other star may remain captured around the black hole as an S-star. HVSs, which were first observed by \cite{2005ApJ...622L..33B}, are so fast (they can reach $\sim 1000$ km\,s$^{-1}$) to overcome the gravitational potential well of the Milky Way (MW) and will likely traverse the Galaxy from its centre to the halo. The first HVS discovered was a 3 $M_\odot$ main-sequence, B-type star moving with a Galactic rest frame velocity $>670$ km\,s$^{-1}$, about twice the Galactic escape velocity at its current distance of 100 kpc to the centre \citep{2005ApJ...622L..33B}. About twenty-one, such stars have been observed by the Multiple Mirror Telescope (MMT) telescope, so far. Most of them are late B-type stars within the mass 
range of 2.5-4 $M_\odot$, at distances of 50-100 kpc from the GC \citep{2014ApJ...787...89B}. 

The properties of HVSs detected by the MMT seem to confirm a mechanism of strong acceleration because they are young low-magnitude stars, which should not exist in the halo of the Galaxy where there is no evidence of star formation. Consequently, a likely hypothesis is that such stars have to be formed elsewhere and ejected till those outer regions \citep{2015MNRAS.454.2677C}. Upon the second $Gaia$ Data Release (DR2) on the 25th April 2018 \citep{2016A&A...595A...1G,2018A&A...616A...1G}, the astrometry and photometry for more than $1.3$ billion sources have been delivered \citep{2018MNRAS.479.2789B} and more than 500 candidates have been proposed for HVSs in the Open Fast Star Catalog\footnote{https://faststars.space}. This catalogue includes parallaxes and proper motions \footnote{https://www.cosmos.esa.int/web/gaia/data} which enable to distinguish among Galactic centre HVSs and other high-velocity stars, such as Galactic disc runaway stars and Galactic halo stars \citep{2018ApJ...866...39B}. Using Gaia DR2 proper motions, a significant fraction of B-type HVSs from HVS survey (MMT) \citep{2014ApJ...787...89B,2015ApJ...804...49B} are still found to be consistent with origin from the GC \citep{2018ApJ...866...39B,2019MNRAS.483.2007E}.

Lots of other mechanisms to explain the existence of HVSs, such as a three-body encounter between a single star and a binary black hole (BBH) \citep{2003ApJ...599.1129Y}, have also been suggested.
A similar frame able to explain huge accelerations of stars after a close encounter with an SMBH is that proposed by \citet{2015MNRAS.454.2677C,2016MNRAS.456.2457A}. The idea is that of the interaction of an orbitally decaying globular cluster approaching a massive black hole.
	
On the other hand, within the central parsec of our Galaxy a population of both young and old stars, known as S-stars, are observed revolving very close (< 0.05 pc) around the Galactic CMO \citep[e.g.][]{2009ApJ...692.1075G,2017gillessen...837...30G,2003ApJ...586L.127G} and provide the strongest constraints on the CMO  mass \citep[e.g.][]{2003ApJ...596.1015S}. The S--star cluster is a dynamically relaxed dense cluster of about 40 stars with the magnitude in the range $m_K=14-17$ \citep{2017habibi...847..120H,2019habibi...872L..15H}. The age estimated for the star S2 is about 6.6 Myr and for the rest of the early-type stars is less than 15 Myr whereas for the late-type stars is $\sim$ 3 Gyr.
The young S-stars have also been classified as B-type stars and have randomly oriented highly eccentric orbits \citep[e.g.][]{2005ApJ...620..744G}, compatible with being the former companion of the HVS in the Hills mechanism. The presence of these young stars in the violent region of the GC is a puzzle, since giant molecular clouds, which are the normal sites of star formation in the Galaxy, would be unable to collapse and fragment in the tidal field of the SMBH \citep{1993ApJ...408..496M}. Several models have been proposed to explain the existence of S-stars near the SMBH but none is completely acceptable \citep{2005PhR...419...65A}. One hypothesis is that the S-stars could be old stars that migrated inward and were rejuvenated by mergers due to collisions with other stars, tidal heating, or envelope stripping \citep{lee1987,genzel2003,davies2005}. However, their relatively normal spectra oppose such an exotic history \citep{2009msfp.book...40F}. An alternative scenario assumes the young stars are carried to the GC while bound to an Intermediate-mass black hole (IMBH), consistently with the hypothesis that an IMBH may still be orbiting within the nuclear star cluster (NSC) \citep{2009ApJ...693L..35M}. The infalling star cluster model can also reproduce the peculiar orbits of S-stars near the SMBH, converting an initially thin, co-rotating disc into a nearly isotropic distribution of stars moving on eccentric orbits around the SMBH. \\

Moreover, G--clouds are other mysterious sources revolving closely around Sgr A* in high-eccentric orbits. High-resolution images of the centre of our Milky Way shows that G2 is a faint gigantic dusty object on a highly eccentric orbit around the Galactic CMO \citep{2012Natur.481...51G,2013ApJ...763...78G}. The semimajor axis and eccentricity of $a = 0.042 \pm 0.01$ pc and $e = 0.98 \pm 0.007$ is estimated for the G2 cloud \citep{2015ApJ...798..111P}.  
Analogous to G2, another dusty, ionized, gas cloud of moderate mass, called G1, was reported in the vicinity of Sgr A* \citep{2005A&A...439L...9C,2005ApJ...635.1087G,2015ApJ...798..111P}. The G1 cloud revolves around Sgr A* on a smaller orbit, $a= 0.0144\pm 0.0064$, with a lower eccentricity, $e=0.860\pm 0.050$ \citep{2015ApJ...798..111P}. Both G1 and G2 objects have already passed their pericentre with no apparent sign of disruption or emission of X-ray flares and the objects are still point-like \citep{mapellireview,haggard2014}. This would imply that these sources likely have the same origin \citep{2015ApJ...798..111P}. Anyway, the origin of G1 and G2 is still an enigma; \citet{2015ApJ...806..197M} showed the consistency of the G2 cloud with a planetary embryo, while \citet{2012NatCo...3E1049M} suggested G2 to be a low-mass star with a protoplanetary disc. To explain the origin of their high eccentricity, \citet{2016trani...831...61T} studied the dynamics of hypothetical planets around stars in the CW disc and in the S-star cluster and they found the semi-major axis and eccentricity of planets escaping from the S-star cluster consistent with those of G1 and G2 clouds. Assuming a stellar origin for G-objects, \citet{trani2019} explain their existence through three--body encounters between binaries of the stellar disk and stellar black holes from a dark cusp around SgrA*.
 
Given the above framework, and given that it is nowadays ascertained that most of the stars host more or less populous planetary systems, even when stars are in binary or multiple systems, it is clear the interest of investigating the fate of planets around binary stars after the binary has experienced a strong interaction with an SMBH because this might eventually lead to, as ejecta, a HVS and a bound S-star. 
Such work has been initiated by \citet{2006MNRAS.368..221G} who examined the fate of former binary companions by simulating 600 different binary orbits around Sgr A* with a direct summation $N$-body code. \citet{2010antonini...713...90A} have studied binary-SMBH encounters and the inclination excitation and eccentricity chaos due to Kozai--Lidov \citep{kozai1962,lidov1962} mechanism leading to collisions and mergers of binary stars. Later, \citet{2012MNRAS.423..948G} have extended their previous work performing a series of simulations of binary stars with planets interacting with Sgr A* massive object, interesting although not conclusive because obtained with a code that does not provide sufficient accuracy for close encounters of objects where the mass ratio (SMBH to planet) is $\approx 10^{9}$, and does not account for relativistic effects. On another side, \citet{2017MNRAS.466.1805F} found that the likelihood of finding exoplanets around high-velocity stars by the transit method depends mainly on mean planetary inclinations and eccentricities (increasing with eccentricity and decreasing with inclination).
They computed the probability of a multi-planetary transit and found it in the range $10^{-3} < P < 10^{-1}$ yr$^{-1}$. Their prediction is that in order to spot a transit it is needed to observe $ \sim 10 - 1000$ stars.  The discoveries of giant exoplanets confirm the giant planet-- metallicity correlation i.e. when the metallicity of a star is high the star is anticipated to host a giant planet \citep[e.g.][]{gonzalez1997,santos2003}. Due to this fact, a metal--rich HVS might host a giant planet or has experienced the coalescence of planet into its surface \citep{2012MNRAS.423..948G}. We underline here that while, at the present time, the exploration of a solitary, super fast and distant hypervelocity planet is challenging, gravitational microlensing could be a suitable technique to detect, indirectly, the presence of small exoplanets at significant distances from the Earth.  In addition, we hope that take advantage of microlensing, we could explore planets around HVSs. The discovery of two planetary systems consisting of a Saturn-mass planet orbiting an M-dwarf, which were detected in too faint and short microlensing events \citep{mroz2017} and also the discovery of extremely ultra--short timescale microlensing events that can be attributed to free-foating or wide-orbit planets \citep{mroz2020}, bring hope for detecting solitary HVPs and planets around HVSs in the future.

Our present work attempts to overcome the limitations of previous theoretical/numerical studies \citep{2006MNRAS.368..221G,2012MNRAS.423..948G,2017MNRAS.466.1805F}, aiming to a more precise prediction of generating high velocity stars with planets. Our improvements regard mainly the use of a high precision and regularized code which treats with accuracy the interaction among objects over an enormous mass range (in our case the mass ratio reaches $1:10^9$) and makes use of a post-Newtonian treatment and allows for external potential and dynamical friction.
	
In section 2 we describe the methodology employed in our simulations and the choice of initial conditions of our runs for the interaction of a four-body system (a binary star, where both of the components host a revolving planet) with an SMBH. The results, including those related to star--star and star--planet collisions and mergers are given in section 3. Finally, our summary and discussion are presented in section 4.
%----------------------------------------------------------------------------------------------------------------------------------------------------------

\section{Model and Methods}
\label{modmeth}
Studying the close interaction of stars and planets with the SMBH is a tough numerical task due to the enormous mass ratio involved ($\gtrapprox 10^{9}$). This would make it almost impossible to numerically follow the planet orbits during the close interaction with the SMBH if using standard integration schemes. 
To do it in a proper, reliable way we carry out our simulations using a regularized $N$-body code, the \archain integrator \citep{1999MNRAS.310..745M,2008AJ....135.2398M}, which includes post-Newtonian (PN) corrections up to order 2.5, properly modified to consider an analytic external potential and its dynamical friction \citep{2019MNRAS.483..152A}.
The most updated version of  \archain \citep{chacd19} contains the possibility of identify individual spin for the black holes, together with a treatment of relativistic kick velocities after BH mergers.\footnote{The free (upon citation) version of the code is available at https://sites.google.com/uniroma1.it/astrogroup/hpc-html.}  \\

A proper modelization of the GC environment, is indeed required when the space scales are not too narrow.  Hence, we take into account the local distribution of stars in the form of a regular external potentials which also induces dynamical friction on orbiting objects.\par

To model the Galaxy density profile in spherical symmetry, we use the sum of a  \citet{1993MNRAS.265..250D} and a  \citet{1911MNRAS..71..460P} distribution:
	
\begin{enumerate}
\item %\textbf{Dehnen Model} 
the galactic background is represented with a Dehnen's (or $\gamma$) whose density  profile is
\begin{equation} 
\rho_D(r)=\frac{(3-\gamma)M_D}{4\pi} \frac{a}{r^{\gamma}(r+a)^{4-\gamma}},\
\end{equation}
where $a$ is a length scale, $M_D$ is the total mass,
and $0\leq\gamma<3$ is a parameter to adjust the steepness of the profile. For our model, we choose $M_D=10^{11} M_{\odot}$, $a$=2 kpc and $\gamma=0.1$, like in \cite{2017MNRAS.471..478A};
\item to model the nuclear star cluster (NSC) around the GC, we use a Plummer density profile  
\begin{equation} 
\label{eq:Plum}
\rho_P(r)=\frac{3M_P}{4\pi}\frac{b^2}{(b^2+r^{2})^{5/2}},
\end{equation}
where $b$ is a length scale and $M_p$ is the model mass. Following \citet{schodel2014}, we adopt $M_p = 2.5\pm 0.4 \times 10^7 M\odot$  and $b=4$ pc.  
\end{enumerate}

In this work, we study the orbital evolution of a binary star system which each star has a planet orbiting around it. We change initial conditions of the system in the Galactic central region as modelled above. 
%Both stars and planets are approximated as point--like masses and 
We set the mass of the SMBH in the origin of the reference frame to $M_{\bullet}=4\times 10^{6} M_\odot$.\\
As mentioned in the introduction, the basic idea to originate high- and even hyper-velocity stars is due to \citet{1988Natur.331..687H}; a binary, moving on a low energy ($E$) orbit, approaches an MBH well within its sphere of influence where it experiences a 3-body interaction. If an exchange collision occurs, the ejection velocity, is given approximately by

\begin{equation} 
\label{eq:v_ej}
\begin{split}
  v_{ej}& \approx 1800 \left (\frac{a_*}{0.1 \textrm{AU}}\right)^{-1/2} \left (\frac{m_{bin}}{2 {M}_\odot}\right)^{1/3}
  \left (\frac{{M}_\bullet}{4 \times 10^6 {M}_\odot}\right)^{1/6}  \textrm{km s}^{-1},
\end{split}
\end{equation}
where $a_*$ and $m_{bin}$ are the separation and total mass of the stars in the binary. Note that this velocity is the velocity at infinity of the ejected star in the absence of the Galactic potential \citep{mer13}. Due to the energy conservation for the 3-body (binary star + SMBH) system, if one star reaches a high velocity increasing significantly its energy, the companion would reduce its orbital energy and, eventually, could be trapped in a orbit around the SMBH, becoming an S-star.
	
In our simulations, the stars in the binary systems are assumed as equal--mass stars with $m_*=3 M_\odot$, similar to the first observed HVS \citep{2015ARA&A..53...15B}, revolving around each other on initially circular orbits at various separations, $a_*$, in the range $0.1-0.5$ AU. This way we almost reproduce the set of HVSs/HVPs studied by \citet{2006MNRAS.368..221G,2012MNRAS.423..948G}, which is useful also for the sake of result comparison.
Each star in the binary hosts one planet with mass $m_p=10^{-3}$ $M_\odot$ (i.e. Jupiter-like),  with circular orbit of radius $a_p=0.02$ AU around the host star.
The initial position of centre of mass of the $4$-body system is located $2000$ AU away from the SMBH. Note that this distance ($0.01$ pc) is well within the SMBH influence radius ($\sim 2.5$ pc) but still $\sim 25,000$ times the SMBH Schwarzschild's radius.

The external potential and dynamical friction, together with the PN terms, might have relevance for orbits sinking very close to the Galactic center
(i.e. highly eccentric). Hence we keep both external potential and PN terms up to order 2.5 in the simulation, after checking that the extra cpu time is not significantly enlarged. Furthermore, all the assumed initial conditions were tested to correspond to initially stable orbits for both binary stars and planets around.

To have a relevant effect for the binary--SMBH interaction, it is obviously necessary to place binary stars onto orbits that come close enough to the SMBH. If the tidal disruption radius of a binary is given by \citep{mer13}
\begin{equation} 
\label{eq:r_bin}
r_t\approx a_*\left (\frac{{M}_\bullet}{{m}_*}\right)^{1/3}
\end{equation}
we have to choose an initial transverse speed ($v_{\bot}$) 
\begin{equation} 
\label{eq:v_ini}
v_{\bot}d=\left (\frac{G {M}_\bullet}{{r}_{min}} \right)^{1/2}{r}_{min}
\end{equation}
for the binary so that $ r_{min}\lesssim r_t$, where $r_{min}$ is the minimum distance of the binary's orbit in the initial approach of the binary towards the SMBH and $d$ is the initial distance of the system's centre of mass from the SMBH \citep{2006MNRAS.368..221G}. Therefore, we give to the system centre of mass a transverse (respect to the binary centre of mass-SMBH joining line) initial speed of $v_{\bot}=66.5$ km s$^{-1}$, which is the maximum speed for systems to enter the binary-SMBH tidal radius.

\begin{figure} 
\begin{minipage}{0.5\textwidth}
\subfloat{%
\includegraphics[scale=1.0]{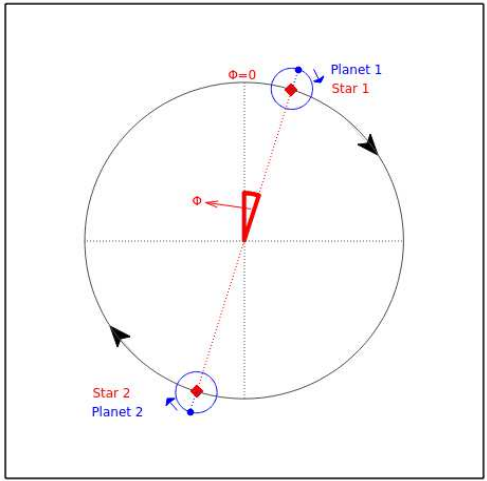}}
\caption{Schematic illustration of the four--body system (two stars in the binary with one planet for each star).} The phase angle ($\phi$) between the position vectors of the two stars (or star-planet).
		
\label{fig:image0}
\end{minipage}\hspace{8em}
\end{figure}

We run simulations at varying: (i) the inclination angle, $i$, of the binary orbital plane respect to its centre of mass orbital plane, choosing the values $0^{\circ}$, $90^{\circ}$ and $180^{\circ}$
(an inclination $0^{\circ}$ means that the four-body system is counter-rotating respect to the centre of mass orbital plane and inclination $180^{\circ}$ is on the opposite), and, (ii), the initial orbital phase angle ($\phi$) of the binary orbit in the whole $0^{\circ}-360^{\circ}$ range at steps of $15^{\circ}$. The phase angle is defined as the initial value of the angle between the 2 stars (and planets) in the assumption that the two planets start moving from positions on the same line joining the two stars, externally to them (see Fig. ~\ref{fig:image0}). In total, we performed 360 simulations which are all extended up to $1600$ years. Such a time corresponds to $\sim 200$ approaches to pericentre by the binary star and to $\sim 975$ times the initial planet orbital period.

Table~\ref{tab:table1} gives the set of values of the initial parameters. Every run is characterized by the  ($a_*, i, \phi$) set of values, because $a_P, m_*$ and $m_P$ are fixed.

\begin{table}
	\centering
	\caption{Initial values for run parameters.Angle $\phi$ varies at $15^\circ$ steps.}
	\label{tab:table1}
	\begin{tabular}{c rrrrrr} % six columns, alignment for each
		\hline\hline           %inserting double-line
		$a_*$ & $a_P$& $m_*$ & $m_P$ & $i$ & $\phi$ \\
		(AU) & (AU)& $M_\odot$ & $M_\odot$ & deg & deg \\[0.5ex]  % [0.5ex] adds vertical space 
		\hline            % inserts single-line
		0.1 & 0.02 & 3 & 0.001 & 0, 90, 180 & 0--360 \\
		0.2 & 0.02 & 3 & 0.001 & 0, 90, 180 & 0--360 \\
		0.3 & 0.02 & 3 & 0.001 & 0, 90, 180 & 0--360 \\
		0.4 & 0.02 & 3 & 0.001 & 0, 90, 180 & 0--360 \\
		0.5 & 0.02 & 3 & 0.001 & 0, 90, 180 & 0--360 \\[1ex] % [1ex] adds vertical space
		
		\hline             %inserts single-line 
	\end{tabular}
\end{table}
	
%-------------------------------------------------------------------------------------------------------------------------------------------

\section{Results}

Our simulations show that after the interaction of the binary system with the SMBH, there are various possible fates for the four-body system components:

\begin{itemize}
\item  one of the stars becomes a high- or even hyper-velocity star, keeping, or losing, its planet; 
\item if one star is expelled at high velocity, the other can be captured around the SMBH, starting to revolve on a precessing, eccentric, orbit (S-star) with or without its planet; 
\item planets can be either driven out and follow unlimited orbits in the Galaxy (hypervelocity planets, HVPs) or revolving independently on highly eccentric orbits around Sgr A* (S-planets), similarly to the S-stars;
\item stars and/or planets can be, also, \textit{swallowed} by the SMBH  that means that the star/planet penetrates within  $3R_{ISCO}$ from the SMBH \footnote{$R_{ISCO}\equiv 6GM_{\bullet}/c^2$ is the radius of the  innermost stable circular orbit (ISCO). For an SMBH of mass $M_{\bullet}=4\times 10^6 M_{\odot}$, R$_{ISCO}$ is  $\approx$ 0.25 AU.};
\item the two stars in the binary can collide with a relative velocity lower than the escape velocity from their surface and merge;
\item  planets around the two stars in the binary could also merge with their host star when the relative velocity upon collision is lower than the escape velocity from the star surface.
\end{itemize}

%%%%%%%%%%%%%%%%%%%%%%%%%%%%%%%%%%%%%%%%%%%%%%%

\begin{figure*} 
\centering

\includegraphics[scale=0.6]{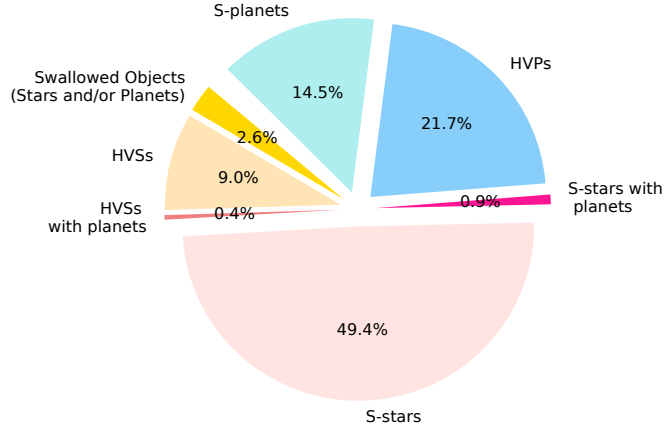}

\caption{The various outcomes of our simulations.}
\label{fig:pie}

\end{figure*}

%%%%%%%%%%%%%%%%%%%%%%%%%%%%%%%%%%%%%%%%%%%%%%%
 
Figure~\ref{fig:pie} represents the results of our simulations in the form of a fractional pie. 

%%%%%%%%%%%%%%%%%%%%%%%%%%%%%%%%%%%%%%%%%%%%%%%%%%%%

\begin{figure*} 
\centering
%\begin{minipage}{0.5\textwidth}
%\subfloat{%
\includegraphics[scale=0.6]{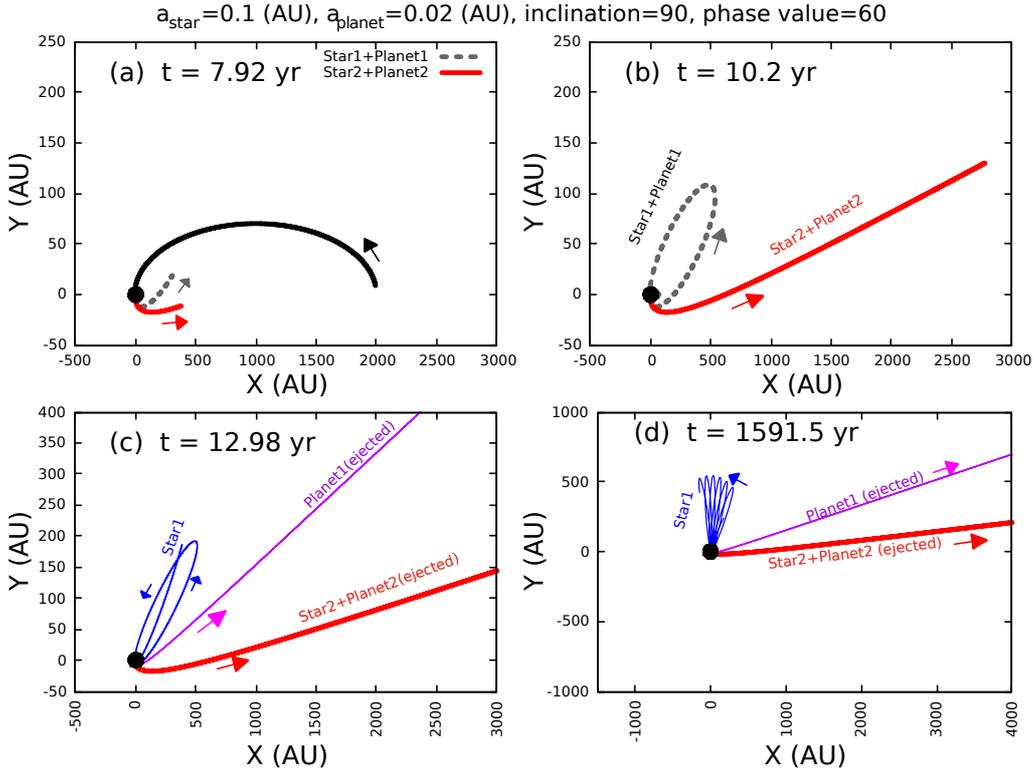}
%}
\caption{Trajectories of the different components of the system ($0.1, 90^\circ, 60^\circ$) as developed up to different times (as labeled). The initial location of the 4--body system is (x,y)=(2000,0) AU and the SMBH is set at the origin. Note that in all the panels the x and y-axes have very different scales to have a sightly zoom of the trajectories.}

\label{fig:image1}
%\end{minipage}
\hspace{8em}
\end{figure*}

\begin{figure*} 
\centering
%\begin{minipage}{0.5\textwidth}
%\subfloat{%
\includegraphics[scale=0.6]{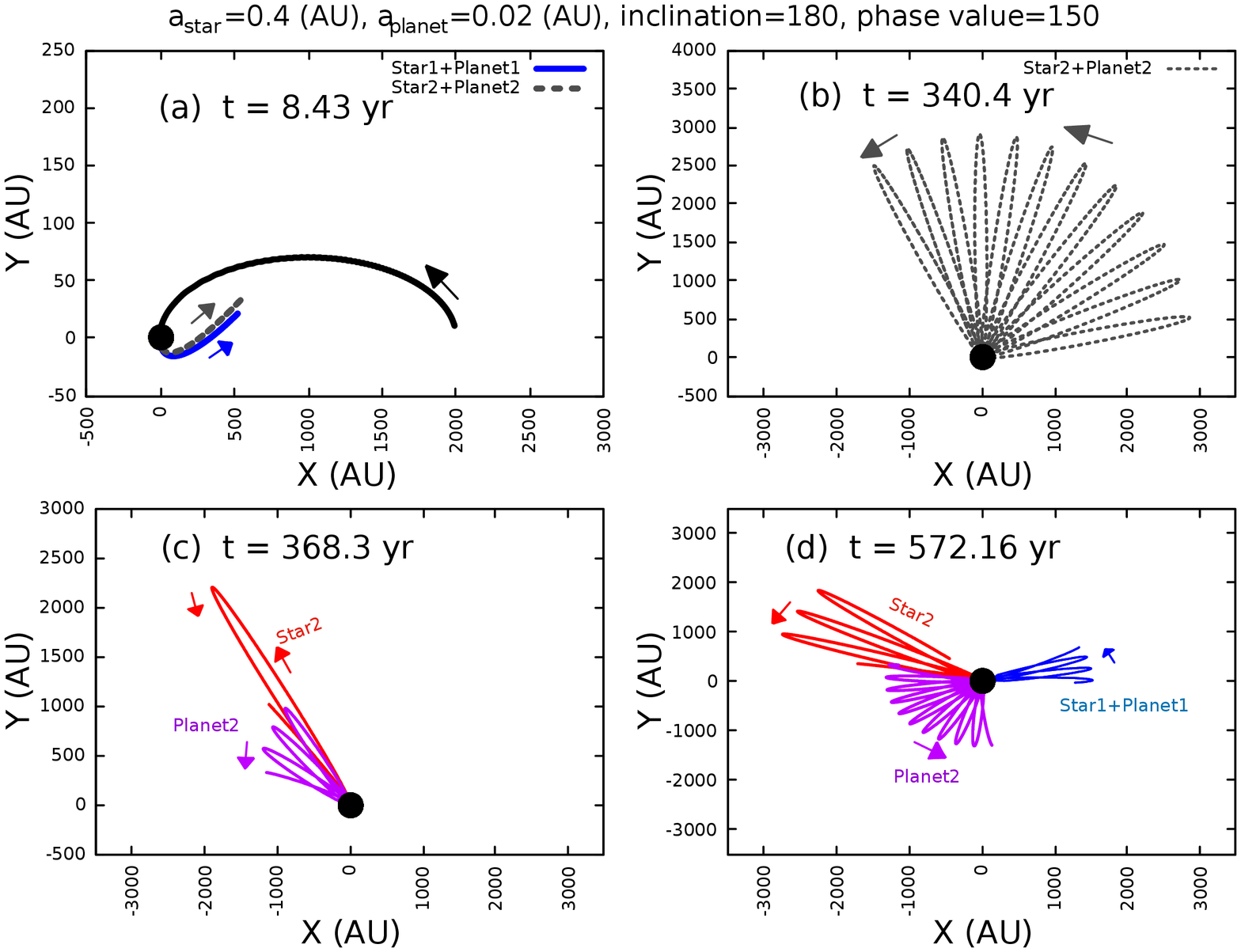}
%}

\caption{% Close approach of the system ($0.4, 180^\circ, 150^\circ$) to Sgr A* taken at different times. 
Trajectories of the different components of the system ($0.4, 180^\circ, 150^\circ$) as developed up to different times (as labeled). The initial location of the 4--body system is (x,y)=(2000,0) AU and the SMBH is set at the origin. Note that, similar to Fig.\ref{fig:image1}, the x and y-axes have very different scales.}

\label{fig:image2}
%\end{minipage}
\hspace{8em}
\end{figure*}

In Figures~\ref{fig:image1} and ~\ref{fig:image2} two examples of our runs are displayed, namely ($0.1, 90^\circ, 60^\circ$) and ($0.4, 180^\circ, 150^\circ$), respectively. These figures show two different outcomes after encounter of the binary system hosting planets with the SMBH in the GC. In Figure~\ref{fig:image1} the initial binary separation is $a_*=0.1$ AU and the binary orbital plane is perpendicular to its centre of mass orbital plane (i=90$^{\circ}$), and the initial phase value of the binary orbit is $\phi=60^{\circ}$. Panel (a) shows the initial trajectory of the four--body system (solid black line) at the time that binary star, which was initially located at (x,y)=(2000,0) AU, is broken up after the first encounter with the SMBH. Panel (b) illustrates the time when HVS hosting planet (red line) is ejected and the other planet is still bound to its host star after one orbital period around the SMBH (dashed grey line). In panel (c), the time at which the star-planet system is separated is shown where the planet is ejected (magenta line) and its host star is orbiting around the SMBH at an eccentric S--star like orbit (blue line). Final fate of the system is demonstrated in panel (d) as single HVP (magenta line), precessing single S-star (blue line) and HVS with planet (red line).

In Figure~\ref{fig:image2} the initial binary separation is larger ($0.4$ AU) and the orbital plane inclination angle is i=180$^{\circ}$; the angular phase of the binary is $\phi=150^{\circ}$. Panel (a) displays the time at which four--body system (solid black line), initially located at (x,y)=(2000,0) AU, is broken up into two star-planet systems. Panel (b) represents the orbital revolution of one star--planet system around the SMBH up to the time the star and its planet are bound to each other. The orbit of the other bound star--planet system is omitted in panels (b) and (c) to avoid confusion. In panel (c), the time which the star--planet system is separated and both star and planet start revolving around the SMBH in individual orbits, is demonstrated. Panel (d) manifests the final fate of all the system's components as precessing single S-star (red line), precessing single S-planet (magenta line) and S-star with planet (precessing blue line).

%%%%%%%%%%%%%%%%%%%%%%%%%%%%%%%%%%%%%%%%%%%%%%%%%%%%

%%%%%%%%%% %%%%%%%%%%%%%%%%%%%%%%%%%%%%%%%%%%%%%%		
		
%%%%%%%%%%%%%%%%%%%%%%%%%%%%%%%%%%%%%%%%%%%%%%

 We underline that an HVS/HVP is the object whose speed at the end of the simulation exceeds the local escape velocity.
In our runs, an HVS/HVP reaches the average distance of $\sim 0.5$ pc from the Galactic Centre. The local escape velocity is given by

\begin{equation}
\begin{split}
v_{esc}(r)&=\sqrt{2U(r)},
\end{split}
\end{equation}
where $U(r)$ is the total potential
\begin{equation}
 U(r)=G\frac{M_{\bullet}}{r}+U_D(r)+U_P(r).  
\end{equation}

 with $U_D(r)$ and $U_P(r)$ representing the  potentials of the Dehnen and Plummer models. When evaluated at $r=0.5$ pc, $v_{esc} \simeq 600$ km s$^{-1}$.

On the other side, we define a star or a planet in the system as an eventual S--star (S--planet) if it keeps orbiting around the central massive object up to the end of simulation.  \\

To investigate the star-star collisions, we consider the  geometrical cross section, $4\pi R_*^2$, where $R_*$ is the radius of the star.

To distinguish the star-planet physical collision from a disruptive encounter, we consider a star-planet collision to occur when the planet approaches a star at a distance less than the maximum between $R_*+R_P$ and the tidal disruption radius.

\begin{equation}
\label{eq:r_tidal}
  r_t\approx R_p\left (\frac{m_*}{m_p}\right)^{1/3}. 
\end{equation}

Indeed the fate of the planet is different if it is physically colliding onto the star surface or entering the tidal radius and undergoing the tidal disruption. Note that for a $3$ M$_\odot$ star and a Jupiter-like planet the two quantities above ($r_t$ and $R_*+R_P$) are same order of magnitude ($\approx 10 R_P$) so it does not make a real difference.

%{\roberto things are like I wrote above. if a planet approaches a star entering the tidal radius its fate is a priori uncertain because it can physically collide onto the star surface just if the disruption takes a time longer than the time needed to collide on the star (time which is $\simeq$ $r_t-(R_\star+R_P)$. In our practical case, due to that $r_t$ is just a little less than $R_\star+R_P$ I would say that the actual fate is that of a physical collision..}

 Moreover, the stars and planets around them will coalesce (merge) when their relative speed upon collision is lower than the escape velocity from the star surface ($\sim$ 500 kms$^{-1}$ for a 3 M$_{\odot}$ star) \citep{ginsburg2007,2010antonini...713...90A}. The results for star--star and star--planet collisions/mergers are shown in Table~\ref{tab:collision} for three different intervals of time. 

As illustrated in Table. \ref{tab:collision}, collisions of the two stars in the binary happen at almost same rate after the first approach to the Sgr A* CMO. In fact, the massive early--type S--stars may be collisional products of lower mass stars if the merger efficiency in high-velocity collisions is high and if angular momentum of the rapidly rotating merger is removed \citep{genzel2003}.
Besides, occurrence of star-planet collisions/mergers is also probable in the central region of the Galaxy due to tidal field of the SMBH. We find that the number of star--planet collision/merger noticeably rises after the first pericentre approach. These star--planet mergers are more significant for the repeated pericentre approaches with the SMBH (see Table. \ref{tab:collision}). Some of the coalesced star-planet systems in our simulations are captured by the SMBH and revolve around it or gain sufficient energy to escape the GC. 

%%%%%%%%%%%%%%%%%%%%%%%%%%%%%%%%%%%%%%%%%%%%%%%%%%

\begin{figure*} 
	\centering
	\includegraphics[scale=0.7]{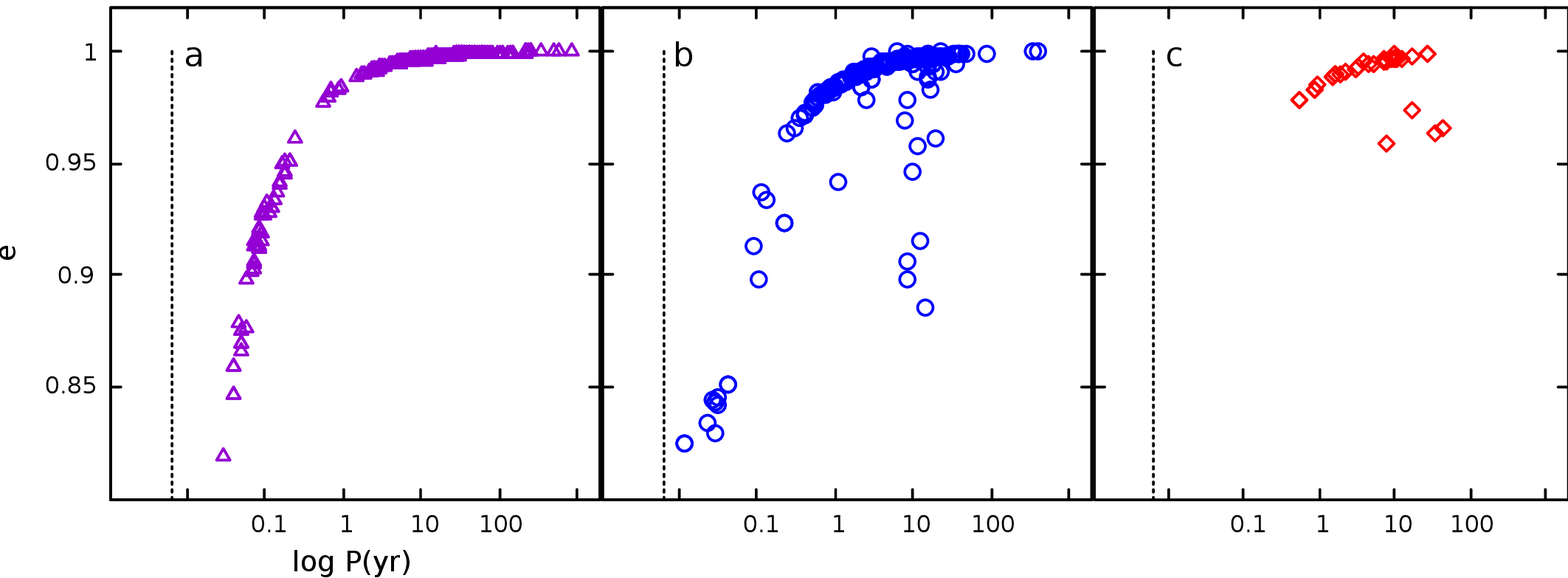}
	\caption{Eccentricity versus period of companion-less S-stars (a), solitary S-planets (b) and S-stars hosting planets (c). The vertical dashed line shows period for an apocentre distance equal to the tidal disruption radius of the binary by the SMBH.}
	\label{ecc}
	\hspace{8em}
\end{figure*}
%%%%%%%%%%%%%%%%%%%%%%%%%%%%%%%%%%%%%%%%%%%%%%%%%%%
%%%%%%%%%%%%%%%%%%%%%%%%%%%%%%%%%%%%%%%%%%%%%%%%%%%%
\begin{table*}
	\centering
	\caption{Percentage factions of collisions and mergers of star--star (top table) and star--planet (bottom table) at three different time intervals. The first pericenter approach of the 4--body system with the SMBH occurs at  $t=8$ yr.}
\label{tab:collision}	
\begin{tabular}{ c c c c c c c c c }
 \hline \hline
 \multicolumn{9}{c}{star--star collisions and mergers fraction (\%)} \\
 \hline \hline\\

$a_*$ & \multicolumn{2}{c}{8 [yr]} & &\multicolumn{2}{c}{8--160 [yr]} &  & \multicolumn{2}{c}{160--1600 [yr]} \\  
\cline{2-3} \cline{5-6} \cline{8-9}
(AU)  & Coll & Merg & & Coll & Merg & & Coll & Merg \\
\hline 
\hline
0.1& 5.55 &   -  && 2.78 & 1.40 && 4.16 & 2.78 \\
0.2& 5.55 & 4.17 && 5.55 & 1.38 && -    & -    \\
0.3& 8.33 & 1.39 && -    & -    && -    & -    \\
0.4& 11.11& 1.37 && -    & -    && -    & -     \\
0.5& 5.55 & 1.37 && -    & -    && -    & -      \\
% \hline
\end{tabular}

\vspace{5mm}

\begin{tabular}{ c c c c c c c c c }
\hline \hline
 \multicolumn{9}{c}{star--planet collisions and mergers fraction (\%)} \\
 \hline \hline\\
$a_*$ & \multicolumn{2}{c}{8 [yr]} & &\multicolumn{2}{c}{8--160 [yr]} & &\multicolumn{2}{c}{160--1600 yr} \\  
\cline{2-3} \cline{5-6} \cline{8-9}
(AU)  & Coll & Merg & & Coll & Merg & & Coll & Merg \\
\hline 
\hline
0.1& 82.64 & 18.06 && - & - && - & - \\
0.2& 13.19 & 3.47 && 37.50 & 6.94 && 6.94 & 2.08 \\
0.3& -     & - && 37.50 & 10.42 && 12.5 & 9.03 \\
0.4& -     & - && 27.78 & 9.03 && 16.67 & 13.19 \\
0.5& 8.33  & 0.69 && 37.50 & 20.83 && 15.28 & 11.8 \\
 \hline
\end{tabular}
\end{table*}
%%%%%%%%%%%%%%%%%%%%%%%%%%%%%%%%%%%%%%%%%%%%%%%%%%

Our simulations generate numbers of solitary S--planets and a small fraction of S--stars with orbiting planets on high-eccentric orbits around Sgr A*. An interesting comparison among different S-cases is provided by the eccentricity vs period plots in Fig.~\ref{ecc}. In Fig. \ref{ecc}, S-stars with planets peaked at higher eccentricity with respect to S-stars with no planets, although it is expected that S-stars at higher eccentricity are less likely to host planets since they have a smaller Hill radius at pericentre. %We checked the possibility of a planet penetrates into the Hill radius of its host S--star using}
%\begin{equation}
%    r_H \approx a(1-e) \sqrt[3]{\frac{m_P}{3m_*}}.
%\end{equation}

 We see that even at high eccentricities, planets' orbits stay within the Hill radius of the host star, thus these planets are safe from being captured by the SMBH. It may not be possible to ascertain whether these planets can survive in long-term evolution around the SMBH. However, the results are consistent with the high eccentricity of the G2 cloud ($e\sim 0.98$) in the GC which is suggested by \citet{murray2012} to be a protoplanetary disc around a low-mass star. Signature of protoplanetary and low-mass protostar outflow candidates within the central pc of Sgr A* has been reported by \citet{yusefzadeh2015a,yusefzadeh2015b,yusefzadeh2017} using radio continuum observations. 
Furthermore, The orbital eccentricity of some of the  S--stars and S--planets obtained in our simulations are compatible with highly-eccentric stars observed in the S--star cluster, such as S14 ($e \sim 0.9761$), S27 ($e \sim 0.952$), S39 ($e \sim 0.9236$), S111 ($e \sim 1.092$) and S175 ($e \sim 0.9867$) \citep{2009ApJ...692.1075G,2017gillessen...837...30G}.

In addition, the fraction of stars thrown out of the GC as HVSs is
$9 \%$, little less than half that of the HVPs ($21.7 \%$).  The dearth of HVSs respect to HVPs is likely due to the small mass of planets with respect to stars. The change in the stars' kinetic energy ($\delta E $) near binary's time of the closest approach to the SMBH is not enough to accelerate stars at high velocities ($\delta E< |E|$). Rather than ejection, stars undergo interactions with the SMBH and are seized around it as S--stars (see Table \ref{tab:table3}). \\

The likelihood of various outcomes depends strongly on the initial configurations of the system. Results significantly depend on the binary orbital plane inclination and the initial separation of the two stars in the binary, as we now discuss.

\begin{enumerate}
\item Orbital plane inclination variation:\  We detect a specific inclination-dependent correlation for the fate of the system after interaction with the SMBH. We find that HVSs/HVPs as well as S--stars/S--planets are produced for all three inclination variations. Nevertheless, we recognise that for both co-rotating ($i=0^{\circ}$) and counter-rotating ($i=180^{\circ}$) coplanar motions the probability to get HVSs with planets is zero. At $i=90^{\circ}$, most of the 4--body systems are separated into two individual star--planet systems at the first pericentre passage to the SMBH hence a star--planet ejection could probably occur. But the star--planet systems at $i=0^{\circ}, 180^{\circ}$ keep orbiting around the SMBH for a few more orbital revolution leading the star--planet systems break-up and no star--planet ejection happens. Moreover, at $i=90^{\circ}$ there is no likelihood of having S--stars with planets around.
In the vicinity of the SMBH, initial high inclination ($i=90^{\circ}$) can bring binaries into Kozai--Lidov regime in which the rate of interactions (such as orbital energy loss) with the SMBH is increased. The Kozai--Lidov mechanism could destabilize the star--planet systems around the SMBH leading to reduction of star--planets distances. Therefore, we detect no S-stars with planets at 90$^{\circ}$ inclined binary planes. Table~\ref{tab:table2} summarizes the probabilities of different outcomes for the three inclinations examined.
In Table \ref{tab:table2}, the second and third columns present the fraction of hypervelocity stars and hypervelocity planets. In the fourth column, the term HVS-P represents the fraction of hypervelocity stars with planets, while the term $S_*-P$ (fifth column) refers to the fraction of S-stars with planets. The fraction of S-stars and S-planets in our simulations are shown in the sixth and seventh columns, respectively. The eighth column indicates the fraction of stars and/or planets \textit{swallowed} by the SMBH.

\item Initial star separation variety: The binary star initial semi-major axis, $a_*$, also contributes significantly to the final outcomes. Table~\ref{tab:table3} illustrates our results for different initial semi-major axes.
The dearth of HVS ejection, causes the lack of HVSs with orbiting planets specially at larger semi-major axis.
For tighter binaries, $a_*=0.1$ and $a_*=0.2$ AU, there is a little probability of having HVSs with planets, which drops to zero for wider initial star separations, that is $a_*=0.3$, $a_*=0.4$ and $a_*=0.5$ AU. As illustrated in Table \ref{tab:table3}, S--stars could keep their associated planets for initial binary separations in the range $0.2 < a_* < 0.5$. This could be due to the fact that when relativistic precession period in the star--planet orbit is smaller than orbital period of the star--planet system around the SMBH, PN precession suppresses Kozai--Lidov oscillations.   
 
This might not be so effective for highly inclined configurations ($i=90^{\circ}$) since Kozai--Lidov plays the dominant role but it could result in the substantial stability of compact star--planet systems at $i=0^{\circ},180^{\circ}$ configurations and production of the S--stars with planets for semi-major axis in the range $0.2<a_*<0.5$ AU. However, at $a_*=0.1$ AU $\sim 83\%$ of star--planet systems collide (see Table \ref{tab:collision}) and the presence of planets around the S--stars comes to be zero (see Table \ref{tab:table3}). 

\end{enumerate}

%----------------------------------------------------------------------------------------------------------

\begin{table*}
\centering
\caption{The fraction of various outcomes for different inclinations.  The quoted values in the table are the Poissonian errors of the mean.
}

\label{tab:table2}
\begin{tabular}{c rrrrrrrr} % eight columns, alignment for each
\hline\hline           %inserting double-line
inclination& HVS&HVP&HVS-P& $S_*-P$&S-star&S-planet& swallowed\\[0.5ex]  
\hline            % inserts single-line
0$^{\circ}$&0.077$\pm$0.05&0.18$\pm$0.08&0.000$\pm$0.00&0.015$\pm$0.01&0.550$\pm$0.24&0.140$\pm$0.06&0.021$\pm$0.01  \\
90$^{\circ}$ & 0.013$\pm$0.05 & 0.28$\pm$0.11 & 0.013$\pm$0.06 & 0.000$\pm$0.00 & 0.392$\pm$0.16 & 0.148$\pm$0.07 & 0.025$\pm$0.01 \\
180$^{\circ}$& 0.060$\pm$0.03 & 0.192$\pm$0.08 & 0.000$\pm$0.00 & 0.012$\pm$0.01 & 0.544$\pm$0.24 & 0.148$\pm$0.06 & 0.031$\pm$0.01\\[1ex]
% [1ex] adds vertical space
\hline             %inserts single-line 
\end{tabular}
\end{table*}   

%----------------------------------------------------------------------------------------------------------
\begin{table*}
\centering
\caption{The fraction of various outcomes for different initial semi-major axis of the binary star.  Columns are the same as Table \ref{tab:table2}. The quoted values in the table are the Poissonian errors of the mean.
}
\label{tab:table3}
\begin{tabular}{c rrrrrrr} % eight columns, alignment for each
\hline\hline           %inserting double-line
$a_*$(AU)& HVS&HVP&HVS-P& $S_*-P$&S-star&S-planet& swallowed\\[0.5ex]   
\hline            % inserts single-line
0.1 & 0.208$\pm$0.10 & 0.236$\pm$0.11 & 0.014$\pm$0.08 & 0.000$\pm$0.00 & 0.361$\pm$0.21 & 0.118$\pm$0.07 & 0.049$\pm$0.02  \\
0.2 & 0.101$\pm$0.04 & 0.229$\pm$0.11 & 0.007$\pm$0.00 & 0.014$\pm$0.03 & 0.437$\pm$0.25 & 0.167$\pm$0.09 & 0.024$\pm$0.01 \\
0.3 & 0.069$\pm$0.04 & 0.215$\pm$0.12 & 0.000$\pm$0.00 & 0.007$\pm$0.00 & 0.514$\pm$0.29 & 0.177$\pm$0.10 & 0.020$\pm$0.01\\
0.4 & 0.034$\pm$0.02 & 0.201$\pm$0.12 & 0.000$\pm$0.00 & 0.049$\pm$0.03 & 0.538$\pm$0.31 & 0.153$\pm$0.08 & 0.024$\pm$0.01\\
0.5 & 0.038$\pm$0.03 & 0.205$\pm$0.12 & 0.000$\pm$0.00 & 0.007$\pm$0.00 & 0.621$\pm$0.35 & 0.115$\pm$0.06 & 0.014$\pm$0.01\\[1ex]% [1ex] adds vertical space
		\hline             %inserts single-line
	\end{tabular}
\end{table*}
%----------------------------------------------------------------------------------------------------------

\section{Concluding discussion and summary}

In this paper, we investigate the dynamics of binary stars, where each star has a planet orbiting around it, as they move toward the Sgr A* massive object in the centre of our Galaxy. This study can be considered also as a deep, quantitative, generalization of the \citet{1988Natur.331..687H} scenario to the presence of planets around a binary star strongly interacting with the Sgr A* massive object. The numerical exploration of such systems interacting with the SMBH is challenging because of the enormous mass ratios (SMBH to planet mass is of order $10^9$). Standard integration techniques fail and high precision, regularized codes are needed. For this reason, only a few works have looked deeply into this kind of problems, so far.

To perform our simulations, we exploit the accuracy and reliability of the regularized $N$-body code \argdf, which is a modified version of the original Mikkola's \archain code (which includes post-Newtonian corrections up to order 2.5) to account for an external potential and its dynamical friction. A total number of 360 simulations, lasting for 1600 years have been performed.

Indeed we find that if the stars/planet are not assumed as point--like masses there is a non--negligible chance for their mutual collisions/mergers in the vicinity of the CMO. In this work, to investigate the likelihood of collisions and mergers in the proximity of the SMBH in the GC, we assign a physical radius to stars/planets as described in details in section \ref{modmeth}.

Our simulations show the intriguing possibility of the existence of hyper-velocity stars borrowing planets around as well as the production of solitary hyper-velocity planets and also creation of planets kept bound to the stars who has become an S-star revolving around the central massive object.

The main conclusions of our work can be summarized as follows:

\begin{itemize}
    \item stars and planets may escape the GC as HVSs or HVPs, respectively, or be captured on very eccentric S--star--like orbits around the SMBH.
    \item the fraction of S--stars are maximal ($\sim 49.4 \%$) while HVS production is $\sim 9.0\%$ because of including PN approximation, stars in the binary do not gain enough energy to escape the GC and consequently stay bound around Sgr A*;
    \item the fraction of S--stars is almost three times that of S--planets ($ 14.5 \%$) because planets are more likely to be accelerated as HVPs than remain around the SMBH;   
    \item HVSs preferentially form in binary systems with small $a_*$, with the frequency of HVPs being almost constant and independent on $a_*$, and always greater than the frequency of HVSs;
    \item we find that the frequency of HVPs should be at least twice that of HVSs;  
    \item the probability to form HVSs is maximal when the initial configuration is co-planar and is minimal when the inclination angle, i, is close to 90$^{\circ}$;    
    \item a minor fraction of HVSs and S--stars keep their planet around; 
    \item the probability to have HVSs with planets around is zero when the initial inclination configuration is co-planar because star--planet systems are more likely to be ejected in the time of the first pericentre to the SMBH when the initial configuration is perpendicular;    
    \item the likelihood of S--stars (and S--stars with planets) production is maximal when the initial configuration is co-planar and is minimal when the inclination angle is 90$^{\circ}$ as a result of Kozai--Lidov mechanism who reduces the star-star and star--planet distances destabilising the binaries allowing bodies collide, merge or tidally break up;
    \item the eccentricity of S-stars hosting planets are large, in the range $0.97-0.98$, similar to that of G2--cloud in the GC;
    \item collisions of the two stars in the binary occur at almost same rate at the time of the first pericentre approach to the Sgr A* black hole;
    \item  the number of star–planet collision/merger are more significant for the repeated pericentre approaches with the SMBH because of the Kozai--Lidov oscillations in the inner star--planet orbits;    
    \item  $\sim 2.6 \%$ of objects are swallowed by the SMBH. Most of the swallow events occur in the time of the first pericentre to Sgr A* and most of the swallowed objects are planets. 

\end{itemize}
  
In conclusion, our simulations deepen previous works by \citet{2006MNRAS.368..221G,2012MNRAS.423..948G} and estimate the ejection and capture probabilities of the four-body system in its interaction with Sgr A* in the general field of the Milky Way. Our results are almost consistent with \citet{2012MNRAS.423..948G} in the frequency of high--velocity star formation for binary separations in the range $0.1,a_*<0.5$ and in the fraction of S--star formation for $a_*=0.1, 0.2$ AU, but we see less high--velocity planet ejection and S--planet creation since we take into account star--planet mergers and also stars/planets who swallowed by the SMBH. In our runs, most of the swallowed objects after the first pericentre approach by the SMBH, are planets indeed. This remarkably reduces the number of planets thus fewer HVPs and S-planets are generated in our simulations compare to \citet{2012MNRAS.423..948G}. Besides, it seems that PN precession together with external potential and its dynamical friction quenches the survival of planets both around their host stars and around Sgr A*. In comparison with \citet{2012MNRAS.423..948G}, our simulations produce less solitary S--planets around Sgr A* and we estimate fewer planets who could tolerate the survival around their host stars close to Sgr A*.\\

\section*{Acknowledgements}
We thank the anonymous referee for constructive comments that helped to improve the presentation of our results. We thank S. Mikkola for useful discussions on the use of \archain code. We are also grateful to M. Arca-Sedda who made available to us the version of \archain \ apt to treat external potential and dynamical friction (\argdf) and who generously assisted us in its use.



\bibliographystyle{mnras}
\bibliography{reference} % if your bibtex file is called reference.bib

%%%%%%%%%%%%%%%%%%%%%%%%%%%%%%%%%%%%%%%%%%%%%%%%%%

% Don't change these lines
\bsp	% typesetting comment
\label{lastpage}
\end{document}